\documentclass[12pt, draftclsnofoot, onecolumn]{IEEEtran}
\usepackage{amssymb}
\usepackage{amsmath}
\usepackage{cite}
\usepackage{url}
\usepackage{xcolor}
\usepackage{cite,graphicx,amsmath,amssymb}
\usepackage{tabularx,colortbl}
\usepackage[english]{babel}
\usepackage{fancyhdr}
\usepackage{subfigure}
\usepackage{threeparttable}
\usepackage{cite}
\usepackage{subfigure}
\usepackage{multirow}
\usepackage{mathrsfs}
\usepackage{balance}
\usepackage{fancyhdr}
\usepackage{mdwmath}
\usepackage{mdwtab}
\usepackage{caption}
\usepackage{amsthm}
\usepackage[xcolor]{changebar}
\usepackage{lipsum}
\usepackage{graphicx}
\usepackage{pifont}
\usepackage{graphicx}

\usepackage{flafter}
\usepackage{comment}
\usepackage{array}
\usepackage{makecell}

\newtheorem{princ}{Principle}

\newtheorem{remark}{Remark}
\newtheorem{theorem}{Theorem}

\newtheorem{lemma}{Lemma}

\newtheorem{corollary}{Corollary}

\newtheorem{proposition}{Proposition}

\makeatletter
\def\ScaleIfNeeded{%
\ifdim\Gin@nat@width>\linewidth \linewidth \else \Gin@nat@width
\fi } \makeatother

\begin{document}

\title{\textcolor[rgb]{0.00,0.00,1.00}{MIMO Assisted Networks Relying on Intelligent Reflective Surfaces}}
\author{Tianwei Hou,~\IEEEmembership{Graduate Student Member,~IEEE,}
        Yuanwei Liu,~\IEEEmembership{Senior Member,~IEEE,}
        Zhengyu Song,
        Xin Sun,
        Yue Chen,~\IEEEmembership{Senior Member,~IEEE,}
        and Lajos Hanzo,~\IEEEmembership{Fellow,~IEEE}
}


\maketitle

\begin{abstract}
\cbstart \textcolor[rgb]{0.00,0.00,1.00}{ Intelligent reflective surfaces (IRSs) are invoked for improving both spectral efficiency (SE) and energy efficiency (EE). Specifically, an IRS-aided multiple-input multiple-output network is considered, where the performance of randomly roaming users is analyzed by utilizing stochastic geometry tools. As such, to distinguish the superposed signals at each user, the passive beamforming weight at the IRSs and detection weight vectors at the users are jointly designed. As a benefit, by adopting a zero-forcing-based design, the intra-cell interference imposed by the IRS can be suppressed. In order to evaluate the performance of the proposed network, we first derive the approximated channel statistics in the high signal-to-noise-ratio (SNR) regime. Then, we derive the closed-form expressions both for the outage probability and for the ergodic rate of users. Both the high-SNR slopes of ergodic rate and the diversity orders of outage probability are derived for gleaning further insights. The network's SE and EE are also derived.} Our numerical results are provided to confirm that: i) the high-SNR slope of the proposed network is one; ii) the SE and EE can be significantly enhanced by increasing the number of IRS elements.
\cbend

\end{abstract}

\begin{IEEEkeywords}
Intelligent reflective surfaces (IRS), multiple-input multiple-output, passive beamforming, stochastic geometry.
\end{IEEEkeywords}

\section{Introduction}
In next-generation (NG) networks, many sophisticated wireless technologies have been proposed, i.e., massive multiple-input multiple-output (MIMO) and intelligent reflective surface (IRS). In the 5G new radio (NR) standard, reaching out beyond 6GHz, the coverage area is significantly reduced~\cite{5G_NR,5G_NR_2}. Since high-frequency signals are sensitive to blockage effects~\cite{obstacles_5GNR} of trees and buildings, the cost-effective technique of IRS\footnote{ \textcolor[rgb]{0.00,0.00,1.00}{ Also known as intelligent reflecting surfaces (IRSs) and large intelligent surfaces (LISs).}}-assisted wireless networks has been proposed~\cite{LIS_compare_relay,LIS_smart}.
\cbstart \textcolor[rgb]{0.00,0.00,1.00}{ An IRS-aided system relies on a large number of reflective elements, each of which can adjust the phase shifts and possibly the amplitude of the incident signals. Hence, an IRS-aided network can be integrated into the present infrastructures~\cite{data_LIS_hu,position_LIS_hu}.}
By aligning signals reflected by the IRS elements, the received signal for both the base station (BS) and users can be constructively raised or mitigated by appropriately adjusting the global CSI~\cite{Hou_Single_UAV,Back_scatter,reconfig_meta_surf_1,reconfig_meta_surf_2}. Hence, IRS-aided networks have received considerable attention as a benefit of their high energy efficiency (EE). \cbend

Because of the revolutionary concept of IRS networks~\cite{Zhao_LIS_maga}, many potential applications were proposed~\cite{lh1,lh2,lh3}, with special emphasis on their physical layer security~\cite{PLS_LIS_ZhangRui}, cell edge enhancement~\cite{LIS_magazine_multi-scenarios}, device-to-device (D2D) communications and IRS-aided simultaneous wireless information and power transfer (SWIPT)~\cite{Swipt_LIS_ZhangRui}.
\cbstart \textcolor[rgb]{0.00,0.00,1.00}{ Previous research contribution illustrates that IRS networks are able to perform better than the relay-aided networks in the case that the number of IRS elements is high enough.} \cbend
\textcolor[rgb]{0.00,0.00,1.00}{Naturally, the BS-users link is critical, and hence the impact of fading environments of IRS networks can enhance the received signals at the desired users or mitigate the interference~\cite{MISO_with_directlink}.} The bit error ratio was evaluated in~\cite{LIS_perform_Anal} in the case of Rayleigh fading.
\cbstart There is likely to be a line-of-sight (LoS) link between the BS and IRSs. Therefore, the ergodic rates were estimated in~\cite{LIS_rice_lower_bound_rate}, where Rician fading environments were applied. \cbend
The asymptotic transmission rate of IRS networks was evaluated in~\cite{LIS_channel_hardening}, where the impact of IRSs on the channel hardening was considered in Rician fading channels.
\cbstart However, since the IRS elements are produced by multiple diodes, the phase shifts of IRS elements are discrete in practice for an IRS-aided network~\cite{Dai_journal_2bit_RIS,TVT2,TVT3,ZhangRui_MISO_beams_discrete_2}. The signal-to-interference-plus-noise-ratio (SINR) was optimized in an IRS-aided network for finite-resolution phase shifters~\cite{MISO_max_SINR}. \cbend
The associated energy consumption model was proposed in~\cite{glob_energy_model,energy_model_LIS}, where the EE of the proposed networks was optimized. \textcolor[rgb]{0.00,0.00,1.00}{Then the power efficiency was optimized in a MISO-aided non-orthogonal multiple access network, and the proposed network outperforms the classic zero-forcing-based active beamforming~\cite{NOMA_RIS_MISO}.}
\textcolor[rgb]{0.00,0.00,1.00}{By contrast, due to the obstacles, the direct link between the BS and users is not available~\cite{RIS_low_complexity}. Hence, the users located in coverage-holes can be served by IRSs.} \textcolor[rgb]{0.00,0.00,1.00}{The blockage effect in the IRS-aided mmWave networks was considered in~\cite{mmWav_blockage}, where the signals were enhanced both for static and moving blockages.} Since IRSs are passive infrastructures, the channel estimation of the BS-IRS-user links becomes an attractive issue~\cite{Channel_estimation_1,Zhou_RIS_precoding,TVT4}.

However, there is a paucity of literature on the impact of user-locations on the attainable performance. Stochastic geometry (SG) constitutes an efficient mathematical tool for capturing the topological randomness of networks~\cite{Stochastic_haen,Stochastic_Geo}. The users were assumed to be randomly located according to a binomial point process (BPP) in~\cite{stocha_Haen_2009}, and the impact of location randomness in cellular networks was evaluated in~\cite{no_shadowing}. The analytical results indicated that the system performance trends were not gravely affected by shadowing. \textcolor[rgb]{0.00,0.00,1.00}{The environmental objects are modeled by a fixed line process~\cite{reconfig_meta_surf_2}, where the results provided the probability that a randomly distributed object is coated with an IRS. Then, in order to study the effect of deploying IRSs in cellular networks, the blockages are modeled by a Boolean process in~\cite{stochastic_random}, and the ratio of blind-spots to adequate coverage areas was derived.} However, in order to conceive a practical IRS network, the user-positions have to be taken into account by using SG, which is the objective of this treatise.


\begin{table*}[ht!]\scriptsize
\caption{Contrasting our contributions to the literature, and the objective of the proposed model.}
\label{table:RIS_compare_benchmarks}
\begin{center}
\resizebox{\textwidth}{!}{\begin{tabular}{|l|c|c|c|c|c|c|c|}
\hline
& \cite{LIS_compare_relay} & ~\cite{LIS_smart} &  \cite{energy_model_LIS}  &  \cite{renzo_RIS_relay} & \cite{Naka_multiple} & \cite{stochastic_geometry_CoMP_Lyu} & Proposed Model \\
\hline
IRS-aided network             & \checkmark & \checkmark  & \checkmark & \checkmark & \checkmark & \checkmark & \checkmark \\
\hline
HD relaying under AF protocol &            &            & \checkmark &            &             &            & \checkmark \\
\hline
HD relaying under DF protocol & \checkmark &            & \checkmark &            &             &            & \checkmark \\
\hline
Approximated distribution     &            &            &            &            &             & \checkmark & \checkmark \\
\hline
Exact distribution            &            &            &            &            &  \checkmark &            & \checkmark \\
\hline
Path loss modeling            &            &            & \checkmark &            &             & \checkmark & \checkmark \\
\hline
IRS positioning               &            & \checkmark &            &            &             &            & \checkmark \\
\hline
Energy model                  &            &            &            & \checkmark &             &            & \checkmark \\
\hline
Stochastic geometry           &            &            &            &            &             & \checkmark & \checkmark \\
\hline
\end{tabular}
}\end{center}
\end{table*}

\subsection{Motivations and Contributions}

On the one hand, the previous contributions~\cite{LIS_magazine_multi-scenarios,ZhangRui_MISO_beams_1,MISO_max_SINR,ZhangRui_MISO_beams_discrete_2,energy_model_LIS,MISO_with_directlink,LIS_perform_Anal,LIS_channel_hardening,LIS_rice_lower_bound_rate} have mainly been focussed on IRS-aided multiple-input single-output (MISO)  networks, where only a single antenna is employed by the users. On the other hand, there is a lack of literature on the impact of the locations of multiple users.
Motivated by the potential benefits of the IRS-aided networks, in this article we will develop the first comprehensive downlink (DL) analysis of a MIMO-IRS framework using tools from SG, which is capable of providing the first mathematical model of the spatial randomness of multiple users.
The proposed MIMO-IRS network has to solve three additional issues: i) Having multiple antennas and multiple IRS elements impose interference on the users; ii) LoS fading environments between the BS and IRSs has to be considered; iii) The passive beamforming technique used at the multiple IRS elements also has to be reconsidered.
In this article, we will show that the active beamforming employed at the BS combined with appropriately chosen detection vectors at the users perform well both in IRS and non-IRS scenarios. Table~\ref{table:RIS_compare_benchmarks} provides a contrasting of our contributions to the literature, and the objective of the proposed model.

\textcolor[rgb]{0.00,0.00,1.00}{ Against the above background, }our contributions can be summarized as follows:
\begin{itemize}
  \item We propose an IRS-aided network, where SG is invoked for modelling the location randomness. By applying zero-forcing-based design, the detection vector for the users combined with passive beamforming at the IRSs are conceived. The outage probabilities, ergodic rates, spectral efficiencies (SEs), and EEs are characterized for evaluating the impact of the proposed design.
\textcolor[rgb]{0.00,0.00,1.00}{  \item Explicitly, we first derive the exact distribution of the channel statistics in the high-signal-to-noise-ratio (high-SNR) regime by deriving the Laplace transform of the PDF. Then, we derive closed-form expressions of the outage probability (OP) for the proposed MIMO-IRS network. Both analytical and asymptotic results are derived. Furthermore, the associated diversity orders are obtained based on the OP developed. The results confirm that the diversity order of the proposed network mainly depends on the distribution of the fading between the BS and IRSs.
  \item We then derive the approximated distribution for the proposed IRS-aided MIMO network. The closed-form expressions of the ergodic rate for the proposed MIMO-IRS network are derived. High-SNR slopes are obtained based on the ergodic rate developed. The closed agreement between the approximated and simulation results confirms the accuracy of our low-complexity analytical method.}
  \item The simulation results confirm our analysis, indicating that: 1) the spatial gain improves the system performance; 2) the distribution of the fading between the IRSs and users only have a modest impact on the network performance; 3) the SE and EE can be significantly improved by increasing the number of antennas at the BS or increasing the number of IRS elements.
\end{itemize}

\subsection{Organization and Notations}

In Section \uppercase\expandafter{\romannumeral2}, our MIMO-IRS networks is discussed. In Section \uppercase\expandafter{\romannumeral3}, the analytical results are evaluated  to quantify the performance attained. Our numerical results in Section \uppercase\expandafter{\romannumeral4} verify the accuracy of our analysis, which is followed by our conclusions in Section \uppercase\expandafter{\romannumeral5}. Table~\ref{TABLE OF NOTATINS} lists some of the critical notations used in this article.
${{\bf{H}}^{\rm{T}}}$, ${{\bf{H}}^{\rm{H}}}$, $rank({\bf{H}})$ and $tr(\rm \bf H)$ denote the transpose, conjugate transpose, rank and trace of the matrix $\rm \bf H$. $\mathbb{P}(\cdot)$ and $\mathbb{E}(\cdot)$ denote the probability and expectation, respectively.
The distribution of a 
random variable with mean $x$ and covariance $k$ is denoted by $\mathcal{RV} (x,k)$; and $\sim$ stands for `distributed as'. \textcolor[rgb]{0.00,0.00,1.00}{$\Gamma ({ \cdot })$ represents the Gamma function, ${}_2{F_2}\left( { \cdot , \cdot ; \cdot , \cdot ; \cdot } \right)$ denotes the generalized hypergeometric series~\cite[eq. (9.141)]{Table_of_integrals}, $\Gamma \left( \cdot, \cdot \right)$ is the Gamma distribution, $\gamma \left( { \cdot , \cdot } \right)$ represents the lower incomplete Gamma function, and $G \left( \cdot \right)$ denotes the Meijer-G function~\cite{Table_of_integrals}.}

\begin{table}
\caption{\\ TABLE OF NOTATIONS}
\centering
\begin{tabular}{|l|r|}
\hline
$\alpha$ & Path loss exponent. \\
\hline
$R$ & The radius of the disc. \\
\hline
$R_m$ & The target rate of user $m$.\\
\hline
$t_1$ &Fading parameter between the BS and IRSs.\\
\hline
$t_2$ &Fading parameter between the IRSs and users.\\
\hline
$M$ & The number of antennas at the BS.\\
\hline
$K$ & The number of antennas at users.\\
\hline
$N$ & The amount number of IRSs.\\
\hline
$\rm \bf H$  & The channel matrix between the BS and IRSs.\\
\hline
${\rm \bf G}_m$ & The channel matrix between the IRSs and user $m$.\\
\hline
$\rm \bf \Phi$ & The effective matrix of IRSs.\\
\hline
\end{tabular}
\label{TABLE OF NOTATINS}
\end{table}

\section{System Model}

\begin{figure}[t!]
\centering
\includegraphics[width =3in]{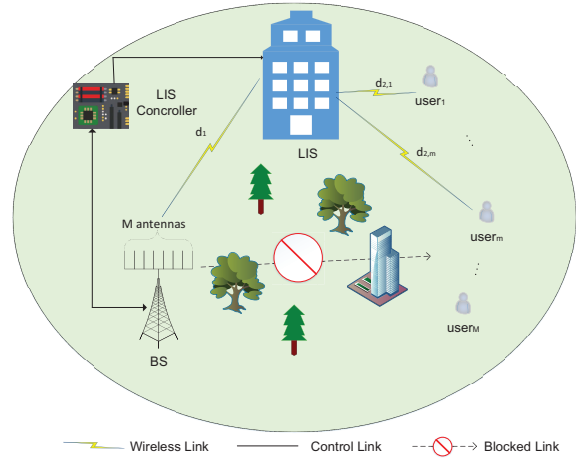}
\caption{\textcolor[rgb]{0.00,0.00,1.00}{ Illustration of a typical IRS assisted wireless transmission model.}}
\label{system_model}
\end{figure}

\textcolor[rgb]{0.00,0.00,1.00}{Consider the IRS-aided MIMO DL, where $M$ transmit antennas (TAs) and $K$ receive antennas (RAs) are equipped at a BS and users, respectively. It is assumed that $N$ IRS elements are co-located at the same building positioned at the center of the disc, and that $N \ge K \ge M$, where the $N$ IRS elements simultaneously serve $M$ users\footnote{\textcolor[rgb]{0.00,0.00,1.00}{Since stochastic geometry tools were invoked, and based on the insights gleaned from~\cite{Hou_multi_UAV}, the object at the origin can be flexibly selected.}}.} By appropriately controlling the phase shifts and amplitude coefficients of the IRS elements, the electromagnetic signal can be beneficially manipulated. Fig.~\ref{system_model} demonstrates the system model of the proposed IRS-aided network. \textcolor[rgb]{0.00,0.00,1.00}{Note that the inter-cell interference significantly affects the network's performance. However, the consideration of a multi-cell setup is beyond the scope of this treatise.}

\subsection{System Description}

\cbstart \textcolor[rgb]{0.00,0.00,1.00}{ Since both the BS and IRSs are infrastructures, hence the locations of the BS and IRSs are fixed. It is assumed that the distances between the BS and IRSs and that between the IRSs and user $m$ are denoted by $d_1$ and $d_{2,m}$, respectively, where $d_{2,m}$ is a random value. } \textcolor[rgb]{0.00,0.00,1.00}{In practice, the distances $d_1$ and $d_{2,m}$ are usually higher than 1 meter for simplifying the analytical results, where the parameter $r_0$ avoids encountering a singularity, when the distance is small~\cite{P_identity_journal_DING}. Generally speaking, $r_0$ is set to 1 meter.} \cbend
\textcolor[rgb]{0.00,0.00,1.00}{ The $M$ users are randomly distributed on a disc $\mathcal{R}^2$ having the radius $R$ according to a BPP, where the direct BS-user transmission links are blocked by potential objects, i.e. trees or buildings~\cite{RIS_low_complexity}.} \cbstart \textcolor[rgb]{0.00,0.00,1.00}{ The path loss of users is expected as product-distance law, which can be expressed as~\cite{renzo_RIS_relay,wave_large_scale}
\begin{equation}\label{large-scale fading}
L_m={(d_1 d_{2,m})^{ - \alpha }},
\end{equation}
where $\alpha$ denotes the path loss exponent.} \cbend

\cbstart Since the LoS links between the BS and IRSs are expected, the small-scale fading matrix is given by \cbend
\begin{equation}\label{channel matrix,BS to LIS}
{\rm \bf H} = \left[ {\begin{array}{*{20}{c}}
{{h_{1,1}}}& \cdots &{{h_{1,M}}}\\
 \vdots & \vdots & \vdots \\
{{h_{N,1}}}& \cdots &{{h_{N,M}}}
\end{array}} \right],
\end{equation}
\textcolor[rgb]{0.00,0.00,1.00}{where ${\rm \bf H}$ is a $N \times M$ matrix whose elements represent the Nakagami fading channel gains, $h_{n,m}$ denotes the channel gain between the $m$-th antenna at the BS and the $n$-th IRS element.}
The probability density function (PDF) of the elements can be expressed as
\begin{equation}\label{channel PDF,eq3}
{f}_1(x) = \frac{{{t_1}^{t_1} {x^{{t_1} - 1}}}}{{\Gamma ({t_1})}}{e^{ - {{{t_1}x}}}},
\end{equation}
where $t_1$ denotes the fading parameter.

\cbstart \textcolor[rgb]{0.00,0.00,1.00}{ Nakagami fading channel is used for evaluating the LoS link between the IRS elements and user $m$, which is given by
\begin{equation}\label{channel matrix,LIS to user}
{\rm \bf G}_{m} = \left[ {\begin{array}{*{20}{c}}
{{g_{m,1,1}}}& \cdots &{{g_{m,1,N}}}\\
 \vdots & \vdots & \vdots \\
{{g_{m,K,1}}}& \cdots &{{g_{m,K,N}}}
\end{array}} \right],
\end{equation}
where ${\rm \bf G}_{m}$ contains $K \times N$ elements associated with the fading parameter $t_2$, $g_{m,k,n}$ denotes the channel gain between the $n$-th IRS element and the $k$-th antenna at user $m$.} \textcolor[rgb]{0.00,0.00,1.00}{Based on the maturity of low-complexity channel estimation algorithms~\cite{MIMO_low_complex_CE}, the global CSI is assumed to be perfectly known at both the IRSs and the users.} 

Hence, the signal received at user $m$ is given by
\begin{equation}\label{received user signal}
{y} =  {\rm \bf G}_{m}  {\rm \bf \Phi}   {\rm \bf H} {\rm \bf P} \sqrt{L_m} + A,
\end{equation}
where \cbstart the reflection-coefficient matrix is a diagonal matrix, which is denoted by ${\rm \bf \Phi}  \buildrel \Delta \over = {\rm{diag}}\left[ {{\beta_1} {\phi _1}, {\beta_2}{\phi _2}, \cdots,{\beta_{N}} {\phi _{N}}} \right]$, where $\beta_n  \in \left( {0,1} \right]$ denotes the amplitude coefficient of IRS elements. The phase shift of IRS element $n$ is denoted by \textcolor[rgb]{0.00,0.00,1.00}{ ${\phi _n} = \exp (j{\theta _n}), j=\sqrt{-1}$, ${\theta _n} \in \left[ {0,2\pi } \right)$, where both the phase shifts and amplitude coefficients are perfectly known at the users~\cite{ZhangRui_MISO_beams_1}.}
 \cbend The additive white Gaussian noise (AWGN) is denoted by $A$ with variance ${\sigma ^2}$. \textcolor[rgb]{0.00,0.00,1.00}{In order to provide access services to multiple users with or without IRSs, the active beamforming weights are set to identity matrix, where ${\rm \bf{P}} = {{\rm \bf{I}}_M}$.} \textcolor[rgb]{0.00,0.00,1.00}{By doing so, the proposed active beamforming weights and detection vectors can be used without any further update in the case that 1) the direct links between the BS and users exist, while no reflected links exist; 2) the reflected links exist, while the direct links between the BS and users do not exist.}

\subsection{Passive Beamforming Design}
\cbstart \textcolor[rgb]{0.00,0.00,1.00}{ Without loss of generality, user $m$ applies a detection vector ${{\rm \bf{v}}_m}$ to the received signal as follows:} \cbend
\begin{equation}\label{received signal after detection}
\begin{aligned}
& {\tilde y}_m = {\rm \bf{v}}_m^{\rm{H}}{{\bf{G}}_{m}}{\bf{\Phi }}{{\bf{H}}}{\rm \bf {P}}\sqrt {{L_m}}  + {\rm \bf{v}}_m^{\rm{H}}{A}\\
&= {\rm \bf{v}}_m^{\rm{H}}{{\bf{G}}_{m}}{\bf{\Phi }}{{\bf{H}}}\sqrt {{L_m}} {{\rm \bf{p}}_m} \\
& + \underbrace {\sum\limits_{i \ne m} {{\rm \bf {v}}_m^{\rm{H}}{{\bf{G}}_{m}}{\bf{\Phi }}{{\bf{H}}}\sqrt {{L_m}} {{\rm \bf{p}}_i}} }_{{\rm{interference}}} + {\rm \bf {v}}_m^{\rm{H}}{A},
\end{aligned}
\end{equation}
where ${\rm \bf{v}}_m$ is a $1 \times K$ vector.

\cbstart \textcolor[rgb]{0.00,0.00,1.00}{ Since the BS-user links are not existed, the channel matrix of the $m$-th user can be transformed into a $K \times M$ matrix, which is given by } \cbend
\begin{equation}\label{effective channels}
{\rm \bf{ H}}_m = {{\rm \bf{G}}_{m}} {\rm \bf \Phi} {{\rm \bf {H}}},
\end{equation}
where the element located at the $k$-th row and $m$-th column represents the signal received at the $k$-th antenna via $N$ IRS elements from the $m$-th antenna of the BS.

\cbstart In order to simultaneously serve all $M$ users by $N$ surfaces, we first define a $K \times N$ complex-valued matrix ${{{\bf{\bar H}}}_m}$ as follows: \cbend
\textcolor[rgb]{0.00,0.00,1.00}{ \begin{equation}\label{complexed vectors of effective channel gain}
\begin{aligned}
&{{{\bf{\bar H}}}_m} \\
& = \left[ {\begin{array}{*{20}{c}}
{{g_{m,1,1}}{h_{1,m}}{\beta _1}{\phi _1}}& \cdots &{{g_{m,1,N}}{h_{N,m}}{\beta _{N}}{\phi _{N}}}\\
 \vdots & \vdots & \vdots \\
{{g_{m,K,1}}{h_{1,m}}{\beta _1}{\phi _1}}& \cdots &{{g_{m,K,N}}{h_{N,m}}{\beta _{N}}{\phi _{N}}}
\end{array}} \right].
\end{aligned}
\end{equation}
} \cbend

\cbstart \textcolor[rgb]{0.00,0.00,1.00}{ Since the superimposed signals for user $m$ cause strong interference, we aim for designing the passive beamforming weights with the goal of mitigating the interference at user $m$. Then, this problem can be formulated by stacking ${{{\bf{\bar H}}}_1}$ to ${{{\bf{\bar H}}}_M}$ as follows:}  \cbend
\textcolor[rgb]{0.00,0.00,1.00}{\begin{equation}\label{find_result}
{\bf{\bar H}} = \left[ \begin{array}{l}
\begin{array}{*{20}{c}}
{{g_{1,1,1}}{h_{1,1}}}& \cdots &{{g_{1,1,N}}{h_{N,1}}}\\
 \vdots & \vdots & \vdots \\
{{g_{1,K,1}}{h_{1,1}}}& \cdots &{{g_{1,K,N}}{h_{N,1}}}\\
{{g_{2,K,1}}{h_{1,2}}}& \cdots &{{g_{2,1,1}}{h_{N,2}}}\\
 \vdots &{}& \vdots \\
{{g_{M,K,1}}{h_{1,M}}}&{}&{{g_{M,K,N}}{h_{N,M}}}
\end{array}
\end{array} \right],
\end{equation}
where ${\bf{\bar H}}$ contains $MK \times N$ elements.}

In order to obtain the passive beamforming weights at the IRS, an object vector is defined as follows:
\begin{equation}\label{maximum channel gain vector}
{\rm \bf S} = {\left[ {\begin{array}{*{20}{c}}
{{s_{1,1}}} \\
{{s_{1,2}}}\\
 \cdots \\
 {{s_{M,K}}}
\end{array}} \right]},
\end{equation}
\cbstart where ${\rm \bf S}$ contains $MK \times 1$ elements, each of which can be obtained by ${{\bf{h}}_m} = {{{\bf{\tilde H}}}_m}{{\bf{1}}_N}$. Hence, for example, the maximum channel gain for the $1$-st antenna at user $m$ can be written as: \cbend
\textcolor[rgb]{0.00,0.00,1.00}{ \begin{equation}\label{maximum achivable element}
\begin{aligned}
{s_{m,1}} &= \left| {\sum\limits_{n = 1}^{N} {{g_{m,1,n}}{h_{n,m}}{\beta _n}{\phi _n}} } \right| \\
 & \le \left| {\sum\limits_{n = 1}^{N} {{g_{m,1,n}}{h_{n,m}}{\beta _n}} } \right|.
 \end{aligned}
\end{equation}}
\textcolor[rgb]{0.00,0.00,1.00}{We assume ${\beta _n} = 1$ for simplicity, and if the reflected signals are co-phased, the maximal achievable channel gain is derived as}
\textcolor[rgb]{0.00,0.00,1.00}{ \begin{equation}\label{maximum channel gain final}
{s_{m,1}} \le \sum\limits_{n = 1}^{N} {} \left| {{g_{m,1,n}}} \right|\left| {{h_{n,m}}} \right|.
\end{equation}}

Hence, we can have the objective function of IRS elements as follows:
\begin{equation}\label{define of the passive beamforming}
\begin{aligned}
&{\Phi _{\rm \bf{v}}} = {{{{\bf{\tilde H}}^{-1}}}}{\rm \bf X}\\
subject \ to \ &  {\theta _1}\cdots {\theta _{N}} \in \left[ {0,2\pi } \right),
\end{aligned}
\end{equation}
\textcolor[rgb]{0.00,0.00,1.00}{where ${\Phi _{\rm{v}}}$ can be decomposed into $ {\left[ {\begin{array}{*{20}{c}}
{{ \bar \beta _1}{\phi _1}},& \cdots , &{{\bar \beta _{N}}{\phi _{N}}} \end{array}} \right]^{\rm{T}}}$, and ${ \bar \beta _n}$ represents the unnormalized amplitude coefficients.}
\textcolor[rgb]{0.00,0.00,1.00}{ Based on the constraint we have ${ \beta _1}, \cdots, { \beta _{N}} \in \left( {0,1} \right] $, whereas the derived unnormalized amplitude coefficients in~\eqref{define of the passive beamforming} may be greater than one, hence we normalize the amplitude coefficients as follows
\begin{equation}\label{normalize amplitude coeff}
{\rm \bf \Phi} = \frac{{\rm \bf \Phi_v}}{{{\beta _{\max }}}},
\end{equation}
where ${{\beta _{\max }}}$ is obtained by finding the maximum amplitude coefficient of ${\Phi _{\rm{v}}}$\footnote{\textcolor[rgb]{0.00,0.00,1.00}{ Note that, since $\beta_{max} \ge 1$, the results derived may not represent the optimized solution.}}.}

\textcolor[rgb]{0.00,0.00,1.00}{ Due to the fact that $rank({{{\bf{\bar H}}}}) \le MK $, therefore no solutions exist for the case of $N < MK $ for passive beamforming. By contrast, for the case of $N > MK $, there exists an infinite number of homogeneous solutions for passive beamforming at the IRS elements, where the constraints of ${\theta _n} \in \left[ {0,2\pi } \right)$ and ${\beta _n} \in \left( {0,1} \right]$, $\forall n = 1, \cdots, N$ can be satisfied.}

\textcolor[rgb]{0.00,0.00,1.00}{By applying the proposed passive beamforming design at the IRSs, the effective channel gains can be derived by ${\rm \bf{v}}_m^{\rm{H}}{{\bf{G}}_{m}}{\bf{\Phi }}{{\bf{H}}}$. Hence, the channel gains for the $m$-th user can be rewritten as
\begin{equation}\label{final effective channel gain}
{{{\rm{ h}}}_m} = \left[ {\begin{array}{*{20}{c}}
{\frac{1}{{{ {{\beta}} _{\rm{max} }}}}\sum\limits_{n = 1}^{N} {} \left| {{g_{m,1,n}}} \right|\left| {{h_{n,m}}} \right|}\\
 \vdots \\
{\frac{1}{{{{\beta} _{\rm{max} }}}}\sum\limits_{n = 1}^{N} {} \left| {{g_{m,K,n}}} \right|\left| {{h_{n,m}}} \right|}
\end{array}} \right].
\end{equation}}

\subsection{Detection Vector Design}
\cbstart The goal of designing detection vectors at users is to ensure that the proposed detection vectors can i) mitigate the intra-cell interference; and ii) perform well both in IRS and non-IRS scenarios, hence the following constraint has to be met: \cbend
\begin{equation}\label{first constraint}
{\rm\bf{v}}_m^{\rm{H}}{\rm \bf{H}}_m{{\rm \bf{p}}_i} = {\rm\bf 0},
\end{equation}
for any $i \ne m$.

\cbstart Since the active beamforming weights are identity matrix, the above constraint in~\eqref{first constraint} can be transformed into \cbend
\begin{equation}\label{second constraint}
{\rm\bf{v}}_m^{\rm{H}}{\rm \bf{ h}}_i = 0,
\end{equation}
\cbstart where ${\rm \bf{h}}_i$ is the $i$-th column of the effective channel matrix ${\rm \bf{ H}}_m$ in~\eqref{effective channels}.
Hence, based on the zero-forcing based design, we first remove the $m$-th column of the channel matrix ${\rm \bf{H}}_m$, we have \cbend
\begin{equation}\label{channel matrix removing effect column}
{\rm \bf{\tilde H}}_m = \left[ {\begin{array}{*{20}{c}}
{\begin{array}{*{20}{c}}
{\begin{array}{*{20}{c}}
{{{{\bf{ h}}}_1}}& \cdots
\end{array}}&{{{{\bf{ h}}}_{m - 1}}}&{{{{\bf{ h}}}_{m + 1}}}
\end{array}}& \cdots &{{{{\bf{ h}}}_M}}
\end{array}} \right].
\end{equation}
Thus, the constraint in~\eqref{second constraint} can be rewritten as follows:
\begin{equation}\label{overall constraint}
{\rm \bf{v}}_m^{\rm \bf {H}}{\rm \bf{\tilde H}}_m = {\rm\bf 0},
\end{equation}
where ${\rm \bf{\bar H}}_m$ contains $K \times (M-1)$ elements.

Hence, we can obtain the detection vector of the $m$-th user from the null space of ${\rm \bf{\tilde H}}_m$, which can be written as
\begin{equation}\label{detection vector}
{{\rm \bf{v}}_m} =  {{\rm \bf{T}}_m}{{\rm \bf{x}}_m},
\end{equation}
where ${{\rm \bf{T}}_m}$ is derived by the left singular vectors of ${\rm \bf{\tilde H}}$. We then use the classic maximal ratio combining (MRC) technique, hence ${{\rm \bf{x}}_m}$ is given by
\begin{equation}\label{X by MRC}
{{\rm \bf{x}}_m} = \frac{{{\rm \bf{T}}_m^{\rm {H}}{{{\rm \bf{h}}}_m}}}{{\left| {{\rm \bf{T}}_m^{\rm{H}}{{{\rm \bf{ h}}}_m}} \right|}}.
\end{equation}
\textcolor[rgb]{0.00,0.00,1.00}{To ensure the existence of the left singular vector in~\eqref{channel matrix removing effect column}, the number of TAs has to be smaller than the number of RAs, i.e. $K \ge M$. Otherwise, the solution does not exist.} \textcolor[rgb]{0.00,0.00,1.00}{Note that more powerful, but more complex active beamformers can be designed for reducing the number of RAs, however, this is beyond the scope of this treatise.}

By using the above active beamforming and DL user-detection vectors, the expectation of the transmit signal power from the BS obeys the maximum transmit power constraint:
\begin{equation}\label{tranmit power constraint}
\mathbb{E} \left\{ {{{\left| {\rm \bf{Ps}} \right|}^2}} \right\} = {\rm{tr}}\left( {{ {\rm \bf P}^H} {\rm \bf P}} \right)p_b,
\end{equation}
where $p_b$ denotes the transmit power of the BS for user $m$. 

\textcolor[rgb]{0.00,0.00,1.00}{ Based on the active beamforming and DL user-detection vectors designed, the SNR of user $m$ after detection can be written as:
\begin{equation}\label{SINR after detection}
SN{R_m} = \frac{{{{\left| {{\rm \bf {v}}_m^{\rm{H}}{{{\bf{ h}}}_m}} \right|}^2}(d_1 d_{2,m})^{ - {\alpha}}{p_b}}}{{{{\left| {{\rm \bf{v}}_m^{\rm{H}}} \right|}^2}{\sigma ^2}}},
\end{equation}
where $\sigma ^2$ denotes the AWGN power.}

\cbstart Since the detection vectors are normalized of user $m$, some further relevant observations are given by: \cbend
\begin{equation}\label{detection vector power 2}
{\left| {{\rm \bf{v}}_m^{\rm{H}}} \right|^2} = {\left( {\frac{{\left| {{{\rm \bf{T}}_m}{\rm \bf{T}}_m^{\rm{H}}} \right|}}{{\left| {{\rm \bf{T}}_m^{\rm{H}}} \right|}}} \right)^2} =  {\left| {{\rm \bf{T}}_m^{\rm{H}}} \right|^2},
\end{equation}
and
\begin{equation}\label{channel}
{\left| {{\rm \bf{v}}_m^{\rm{H}}{{{\bf{ h}}}_m}} \right|^2} = {\left( {\frac{{\left| {{{\rm \bf{T}}_m}{\rm \bf{T}}_m^{\rm{H}}{\rm \bf{ h}}_m^2} \right|}}{{\left| {{\rm\bf{T}}_m^{\rm{H}}{{{\rm \bf{ h}}}_m}} \right|}}} \right)^2} =  {\left| {{\rm \bf{T}}_m^{\rm{H}}{\rm \bf{ h}}_m^{}} \right|^2}.
\end{equation}

\textcolor[rgb]{0.00,0.00,1.00}{ Based on the effective channel gain given by~\eqref{final effective channel gain} and noting that ${\rm\bf{T}}_m^{\rm{H}}{{\rm \bf{T}}_m} = {{\rm \bf{I}}_{Q}}$ in~\eqref{detection vector power 2} along with the effective antenna gain $Q=K-M+1$, the channel gain of user $m$ can be transformed into~\cite{P_identity_journal_DING}
\begin{equation}\label{final effective gain at m user}
\left | {{{{\rm \bf{h}}}_m}} \right | = \frac{1}{{\beta _{\max }}} \left[ {\begin{array}{*{20}{c}}
{\sum\limits_{n = 1}^{N} {} \left| {{g_{m,1,n}}} \right|\left| {{h_{n,m}}} \right|}\\
 \vdots \\
{\sum\limits_{n = 1}^{N} {} \left| {{g_{m,Q,n}}} \right|\left| {{h_{n,m}}} \right|}
\end{array}} \right].
\end{equation}}

\textcolor[rgb]{0.00,0.00,1.00}{ Thus, the SNR of user $m$ can be expressed as
\begin{equation}\label{final SINR}
SN{R_m} = \frac{{\left | {{{{\rm \bf{ h}}}_m}} \right |^2{(d_1 d_{2,m})}^{-\alpha}{p_b}}}{ {\beta _{\max }^2}{\left\|  {{{\rm{I}}_{Q}}} \right\|_2^2{\sigma ^2}}}.
\end{equation}}

\section{Performance Analysis of MIMO-IRS Network}

\cbstart \textcolor[rgb]{0.00,0.00,1.00}{In this section, we discuss the performance of the proposed MIMO-IRS network. The channel statistics derived for the high-SNR regime, the outage probabilities, ergodic rates, SE and EE are evaluated.} \cbend

\subsection{\textcolor[rgb]{0.00,0.00,1.00}{ Channel Statistics}}
\cbstart \textcolor[rgb]{0.00,0.00,1.00}{ We first derive the effective channel statistics for the high-SNR regime of the proposed MIMO-IRS network.} \cbend

\textcolor[rgb]{0.00,0.00,1.00}{Let us assume that users are distributed according to a BPP, where the IRS elements are co-located at the center of the disc, the probability density functions (PDFs) of the user distances are given by}
\begin{equation}\label{PDF of distance distribution}
{f_d}\left( r \right) = \frac{2r}{{ (R^2-r_0^2)}}, {\rm{if}}~~  r_0<r<R.
\end{equation}

\begin{lemma}\label{lemma2:new state of effective channel gain exact high SNR for OP}
Assuming that the elements in ${{\rm \bf{H}}}$ and ${{\rm \bf{G}}_{m}}$ are i.i.d. along with fading parameters $t_1$ and $t_2$ with ${t_1} \ne {t_2}$\footnote{\textcolor[rgb]{0.00,0.00,1.00}{Since both the BS and the IRSs are part of the infrastructure, whereas the users are randomly located on the ground, the fading environments of the BS-IRS links are usually stronger than that of the IRS-user links.}}, respectively. In the proposed MIMO-IRS network, ${N}$ IRS elements simultaneously serve $M$ users, where $N \ge K \ge M$. The PDF of the effective channel gain at user $m$ in the high-SNR regime is given by
\textcolor[rgb]{0.00,0.00,1.00}{ \begin{equation}\label{New Gamma distribution in Lemma_high-SNR-exact PDF}
{f_{\left | {{{{\rm \bf { h}}}_m}} \right |^2}}(x) = \frac{{{{\tilde m}^{N}}}}{{\Gamma \left( {2{t_s}N} \right)}}{x^{2{t_s}N - 1}}{e^{ - 2\sqrt {{t_s}{t_l}} x}},
\end{equation}
where ${t_s} = \min \left\{ {{t_1},{t_2}} \right\}$, ${t_l} = \max \left\{ {{t_1},{t_2}} \right\}$, and $\tilde m = \frac{{\sqrt \pi  {4^{{t_s} - {t_l} + 1}}{{\left( {{t_s}{t_l}} \right)}^{{t_s}}}\Gamma \left( {2{t_s}} \right)\Gamma \left( {2{t_l} - 2{t_s}} \right)}}{{\Gamma \left( {{t_s}} \right)\Gamma \left( {{t_l}} \right)\Gamma \left( {{t_s} + {t_l} + 0.5} \right)}}$.}
The cumulative density function (CDF) of the effective channel gain in the high-SNR regime is given by
\textcolor[rgb]{0.00,0.00,1.00}{ \begin{equation}\label{New Gamma distribution in Lemma_high-SNR-exact CDF}
\begin{aligned}
{F_{\left | {{{{\rm \bf{h}}}_m}} \right |^2}}(x) &= \frac{{{{\tilde m}^{N}}{{\left( {4{t_s}{t_l}} \right)}^{ - {t_s}N}}}}{{\Gamma \left( {2{t_s}N} \right)}} \\
& \times  \gamma \left( {2{t_s}N,2\sqrt {{t_s}{t_l}} x} \right).
\end{aligned}
\end{equation}}
\begin{proof}
Please refer to Appendix A.
\end{proof}
\end{lemma}

\subsection{Outage Probability}

In this article, the OP of user $m$ is defined by
\begin{equation}\label{Outage Defination}
{P_m} = \mathbb{ P} \left( {{{\log }_2}(1 + SN{R_m}) < {R_m}} \right),
\end{equation}
where ${{R_m}}$ denotes the target rate of user $m$. Then we turn our attention to calculating the OP of user $m$, which is given by the following Theorem.

\begin{theorem}\label{Theorem1:Outage m-th user closed form by hypergeom}
\emph{Assuming that the IRS array is located at the center of the disc, and ${N}$ IRS elements simultaneously serve $M$ users, the closed-form OP expression of user $m$ can be expressed as}
\begin{equation}\label{outage analytical results m-th in theorem1}
\begin{aligned}
{P_{m,l}}& = {\tau _{1}}R^{\alpha {a} + 2}{}_2{F_2}\left( {{a},{a} + {\delta _1};{a} + 1,{a} + {\delta _1} + 1; - {b_l}R^\alpha } \right) \\
&  - {\tau _{1}}r_{0}^{\alpha {a} + 2}{}_2{F_2}\left( {{a},{a} + {\delta _1};{a} + 1,{a} + {\delta _1} + 1; - {b_l}r_{0}^\alpha } \right)  ,
\end{aligned}
\end{equation}
\emph{where we have ${\delta _{m,l}} = \frac{{{\varepsilon _m}\beta _{\max }^2Q{\sigma ^2}}}{{{p_b}}}$, ${\varepsilon _m} = {2^{{R_m}}} - 1$, $a = 2{t_s}N$, $b_l = 2\sqrt {{t_s}{t_l}} {\delta _{m,l}}d_1^\alpha $, ${\delta _1} = \frac{1}{\alpha }$, $\varphi  = \frac{{2{{\tilde m}^{N}}{{\left( {4{t_s}{t_l}} \right)}^{ - {t_s}N}}}}{{\Gamma \left( {2{t_s}N} \right)\left( {{R^2} - r_0^2} \right)}}$, and ${\tau _{1}} = \frac{{\varphi {b_l^a}}}{{a(\alpha a + 2)}}$.}
\begin{proof}
Please refer to Appendix B.
\end{proof}
\end{theorem}

\textcolor[rgb]{0.00,0.00,1.00}{ Since ${\delta _m}$ is a function of the maximum amplitude coefficients, it is hard to obtain analytical insights from~\eqref{outage analytical results m-th in theorem1}.
Thus, we turn our attention to finding the optimized solutions.
In order to provide some fundamental engineering insights, we mainly focus our attention on the optimized scenario in the rest of this article, where $\beta_n=1, \forall n$.
It is worth noting that for the case of $K=M=1$, the amplitude coefficients $\bar \beta_n$ and $\beta_{\rm{max}}$ can all be considered to be one and the maximum signal power can be obtained by appropriately adjusting the phase shifts. Thus, we formulate the upper bound of the approximated OP in the following Theorem.}

\begin{theorem}\label{Theorem2:Outage m-th user closed form by hypergeom in minimum achivable}
\emph{Assuming that the number of IRS elements is high enough, the upper bound of the approximated OP is given by}
\begin{equation}\label{outage analytical results m-th in theorem2 minimum achivable}
\begin{aligned}
{P_{m}}& = {\tau _{1}}R^{\alpha {a} + 2}{}_2{F_2}\left( {{a},{a} + {\delta _1};{a} + 1,{a} + {\delta _1} + 1; - {b}R^\alpha } \right) \\
&  - {\tau _{1}}r_{0}^{\alpha {a} + 2}{}_2{F_2}\left( {{a},{a} + {\delta _1};{a} + 1,{a} + {\delta _1} + 1; - {b}r_{0}^\alpha } \right)  ,
\end{aligned}
\end{equation}
\emph{where we have ${\delta _m} = \frac{{{\varepsilon _m}Q{\sigma ^2}}}{{{p_b}}}$, $b = 2\sqrt {{t_s}{t_l}} {\delta _{m}}d_1^\alpha $, and ${\tau _1} = \frac{{\varphi b_{}^a}}{{a(\alpha a + 2)}}$. }
\begin{proof}
Similar to Appendix~B, the results in~\eqref{outage analytical results m-th in theorem2 minimum achivable} can be readily proved.
\end{proof}
\end{theorem}

It is however quite challenging to directly obtain engineering insights from~\eqref{outage analytical results m-th in theorem2 minimum achivable} due to the Gauss hypergeometric function, hence the asymptotic behavior is analyzed in the high-SNR regime in the case of ${\frac{p_b}{\sigma^2}  \to \infty }$.
\begin{corollary}\label{corollary1:Outage m user asymptotic}
\emph{Assuming that $bR_{\max }^\alpha  < 1$, the upper bound of closed-form asymptotic OP expression of user $m$ is given by}
\begin{equation}\label{asymptotic result m user in corollary1}
\begin{aligned}
{{\bar P}_m} = {\frac{\varphi {b^{n + a}}}{a ({\alpha a + 2})}\sum\limits_{n = 0}^\infty  {{t_{n1}}} \left( {R^{\alpha a + \alpha n + 2} - r_{0 }^{\alpha a + \alpha n + 2}} \right)  },
\end{aligned}
\end{equation}
\emph{where~${t_{n1}} = \frac{{{{(a)}_n}{{(a + {\delta _1})}_n}}}{{{{(a + 1)}_n}{{(a + {\delta _1} + 1)}_n}n!}}$, ${ ({{x}})_n}$ represents the Pochhammer symbol~\cite{Table_of_integrals}, which can be calculated as $\frac{{\Gamma \left( {{x} + n} \right)}}{{\Gamma \left( {{x}} \right)}}$ }.
\begin{proof}
Please refer to Appendix C.
\end{proof}
\end{corollary}

\begin{proposition}\label{proposition1: m diversity order}
\emph{From \textbf{Corollary~\ref{corollary1:Outage m user asymptotic}}, we can determine the diversity order by using the asymptotic results, and the diversity order of user $m$ in the proposed MIMO-IRS network is given by}
\begin{equation}\label{diversity order}
{d_m} =  - \mathop {\lim }\limits_{\frac{{{P_b}}}{{{\sigma ^2}}} \to \infty } \frac{{\log {{\bar P}_m}}}{{\log \frac{{{P_b}}}{{{\sigma ^2}}}}} \approx a = 2{t_s}N.
\end{equation}
\textcolor[rgb]{0.00,0.00,1.00}{ \begin{proof}
Based on the results derived in~\eqref{asymptotic result m user in corollary1}, and on the definition of diversity order, we have
\begin{equation}\label{diversity proof1}
{d_m} \approx  - \mathop {\lim }\limits_{\frac{{{P_b}}}{{{\sigma ^2}}} \to \infty } \frac{{\log \sum\limits_{n = 0}^\infty  {\left( {\frac{{{\sigma ^2}}}{{{P_b}}}} \right)_{}^{n + a}} }}{{\log \frac{{{P_b}}}{{{\sigma ^2}}}}}.
\end{equation}
Due to the fact that $\frac{{{P_b}}}{{{\sigma ^2}}} \rightarrow \infty$ and ${\delta _m} \rightarrow 0$, and because the maximal component can be obtained in the case of $n=0$ in~\eqref{diversity proof1}, we have
\begin{equation}\label{diversity proof2}
{d_m}= n+a = a.
\end{equation}
The proof is complete.
\end{proof}}
\end{proposition}

\begin{remark}\label{remark1:impact of fading environment}
\textcolor[rgb]{0.00,0.00,1.00}{ The results in~\eqref{diversity order} demonstrate that both the fading links between the IRS elements and the users as well as that between the BS and IRS elements significantly affect the OP. It is also worth noting that for the case of $Q \to \infty$, the diversity order of the proposed network approaches $2{t_s}N$.}
\end{remark}

\begin{remark}\label{remark2:impact of antenna numbers}
\textcolor[rgb]{0.00,0.00,1.00}{ For the case of $M=1$, the minimal diversity order of the proposed network is $2{t_s}N$, which indicates that having more IRS elements reduces the OP. It is also worth noting that the minimum diversity order of the proposed MIMO-IRS network is $2$ in the case of ${t_1}= N= M=K=1$ and $t_2 >1$, i.e. ${\rm{min}}({d_{m}}) = 2$.}
\end{remark}

It is worth mentioning that a LoS link is expected between the BS and IRSs, and the asymptotic result mainly depends on the $0$-th ordered element in~\eqref{asymptotic result m user in corollary1}. Thus, we provide a basic numerical insight using the following Corollary.
\begin{corollary}\label{corollary4:Outage m user asymptotic MIMO-C-LIS special value}
\emph{Assuming that $M=K=t_2=1$, $t_1 \to \infty$, and $r_0=0$, the $0$-th ordered element in terms of the asymptotic OP of user $m$ is given by}
\textcolor[rgb]{0.00,0.00,1.00}{ \begin{equation}\label{asymptotic result m user in corollary2 MIMO-C-LIS  special velue}
\begin{aligned}
{{\bar P}_{m,0}} = {\frac{2 {\varepsilon _{m}^{N}}  {(d_1 R)}^{N \alpha}}{({N \alpha+ 2}){N!}}}.
\end{aligned}
\end{equation}}
\end{corollary}

\begin{remark}\label{remark-pre:numerical result MIMO-C_LIS}
The results in~\eqref{asymptotic result m user in corollary2 MIMO-C-LIS  special velue} indicate that the diversity order is $N$ in the case of the proposed MIMO-IRS network for $M=K=t_2=1$, $t_1 \to \infty$.
\end{remark}

\begin{remark}\label{remark3:K=M=1,N=3,m2=1,m1}
We also provide a basic engineering insight relying on the proposed network, where $M=K=1$. It is reasonable to assume that no LoS link exists between the IRSs and the user, where the Nakagami fading parameter is $t_2=1$. Thus, based on the results in~\eqref{diversity order}, we can readily infer that the diversity order approaches ${d_m}=2N$.
\end{remark}

\subsection{Ergodic Rate}

\cbstart Since the ergodic rate is able to evaluate the network's SE and EE, we then analyze the ergodic rate of the typical user $m$. In order to obtain a fast and tractable ergodic rate, we first provide the approximated channel gain ${{{\rm \bf{\hat h}}}_m}$ of our MIMO-IRS network.\cbend
\begin{lemma}\label{lemma2:new state of effective channel gain exact}
\textcolor[rgb]{0.00,0.00,1.00}{ Based on the insights of~\cite{Backscatter_naka}, the classic Gamma distribution is assumed.} The approximated distribution of the channel gain at the $m$-th user is given by
\textcolor[rgb]{0.00,0.00,1.00}{ \begin{equation}\label{New Gamma distribution in Lemma}
\left | {{{{\rm \bf{h}}}_m}} \right |^2 \sim \Gamma \left( {\frac{{NQ}}{{{t_h}}},{t_h}} \right),
\end{equation}
where ${t_h} = \frac{{\left( {1 + {t_1} + Q{t_2}} \right)}}{{{t_1}{t_2}}}$.}
\begin{proof}
Please refer to Appendix D.
\end{proof}
\end{lemma}

\cbstart
We then analyze the exact ergodic rate expressions by Meijer-G functions in the following Theorem.
\cbend
\begin{theorem}\label{theorem3:ergodic rate m-th user}
\emph{The lower bound of the achievable ergodic rate of user $m$ can be expressed in closed-form as follows:}
\begin{equation}\label{asympto m-th erogodic rate in corollary4}
\begin{aligned}
{\bar R_{m}} &= \frac{\varphi {c^{ - {\delta _2}}} }{2{\ln (2)}}{\rm{G}}
\begin{tiny}
\begin{array}{*{20}{c}}
{3,1}\\
{2,3}
\end{array}\end{tiny} \left( {\left. {\begin{array}{*{20}{c}}
{{\delta _2},1}\\
{{\delta _2},0,a + {\delta _2}}
\end{array}} \right|cr_{0 }^\alpha } \right)
 \\
& + \frac{\varphi R^2}{2{\ln (2)}} {\rm{G}}
\begin{tiny}
\begin{array}{*{20}{c}}
{3,1}\\
{2,3}
\end{array}
\end{tiny}
\left( {\left. {\begin{array}{*{20}{c}}
{0,1}\\
{0,0,a}
\end{array}} \right|{c}R^\alpha } \right)\\
& - \frac{\varphi r_0^2}{2{\ln (2)}} {\rm{G}}
\begin{tiny}
\begin{array}{*{20}{c}}
{3,1}\\
{2,3}
\end{array}
\end{tiny}
\left( {\left. {\begin{array}{*{20}{c}}
{0,1}\\
{0,0,a}
\end{array}} \right|{c}r_{0 }^\alpha } \right) \\
&  - \frac{\varphi {c^{ - {\delta _2}}} }{2{\ln (2)}}{\rm{G}}
\begin{tiny}
\begin{array}{*{20}{c}}
{3,1}\\
{2,3}
\end{array}
\end{tiny}
\left( {\left. {\begin{array}{*{20}{c}}
{{\delta _2},1}\\
{{\delta _2},0,a + {\delta _2}}
\end{array}} \right|{c}R^\alpha } \right),
\end{aligned}
\end{equation}
where ${c} = \frac{{Q{\sigma ^2}}}{{{t_h}{p_b}}}$.
\begin{proof}
Please refer to Appendix D.
\end{proof}
\end{theorem}

\cbstart To illustrate the increasing slope of the ergodic rate, the high-SNR slope is worth estimating, which can be expressed as\cbend
\begin{equation}\label{High SNR slope of m}
Z = -\mathop {\lim }\limits_{ \frac{p_b}{\sigma^2} \to \infty } \frac{{{\bar R_{m}}}}{{{{\log }_2}\left( {1 + \frac{p_b}{\sigma^2}} \right)}}.
\end{equation}

\begin{proposition}\label{proposition4: m high SNR slopes in massive LIS}
\emph{ By substituting~\eqref{asympto m-th erogodic rate in corollary4} into~\eqref{High SNR slope of m}, the high-SNR slope of the $m$-th user can be derived as}
\begin{equation}\label{high SNR slope in massive LIS}
Z = 1.
\end{equation}
\end{proposition}

\begin{remark}\label{remark6:impact of ergodic rate in massive LIS}
\cbstart \eqref{high SNR slope in massive LIS} indicates that the number of IRS elements has no impact on the high-SNR slope, which is one. \cbend
\end{remark}

\begin{remark}\label{remark6:impact of ergodic rate}
\eqref{high SNR slope in massive LIS} indicates that the slope of the ergodic rate is one, which is not affected by the number of TAs/RAs and the fading environments.
\end{remark}

\subsection{SE and EE}

\cbstart Finally, the SE and EE are obtained in the following two Propositions. \cbend
\begin{proposition}\label{proposition5: spectrum efficiency}
\emph{\cbstart Based on the derived ergodic rates in~\eqref{asympto m-th erogodic rate in corollary4}, an tractable SE can be derived as \cbend }
\begin{equation}\label{spectrum efficiency}
SE = \sum\limits_{m = 1}^M {{\bar R_{m}}} .
\end{equation}
\end{proposition}

\cbstart Since EE is a critical performance indicator, the power consumption can be modelled as~\cite{EE_model_massive_MIMO} \cbend
\begin{equation}\label{total energy massive LIS}
{P_{e}} = {P_{{\rm{B,s}}}} + M P_{\rm{U}} +{M}p_b{{{\varepsilon _b}}} + {N}{P_L},
\end{equation}
\cbstart \textcolor[rgb]{0.00,0.00,1.00}{ where the power consumption and the efficiency of the power amplifier at the BS are denoted by ${P_{{\rm{B,s}}}}$ and ${{\varepsilon _b}}$, respectively. The power consumption of each user and each IRS element are denoted by $P_{\rm{U}}$ and ${P_L}$, respectively.} \cbend

\begin{proposition}\label{proposition7: energy efficiency massive LIS}
\emph{The EE of the proposed MIMO-IRS network can be given by }
\begin{equation}\label{energy efficiency massive LIS}
{\Theta _{EE}} = \frac{SE}{{{P_{e}}}},
\end{equation}
where $S$ and $P_{e}$ are obtained from~\eqref{spectrum efficiency} and~\eqref{energy efficiency massive LIS}, respectively.
\end{proposition}

\section{Numerical Results}

\cbstart We provide numerical results for the performance evaluation of the proposed MIMO-IRS network in this section.
Monte Carlo simulations are provided to confirm the correctness of the analytical results.
The transmit bandwidth is set to $D=100$ MHz, and the power of the AWGN is set to $\sigma^2= -174+ 10 {\rm{log}}_{10}(D)$ dBm.
Nakagami fading parameter $t_1 = 1$ and $t_1 >1$ represent LoS and Non-LoS (NLoS) links, respectively.  \cbend
The path loss exponent is set to $\alpha=3$, and the power attenuation at the reference distance is set to~-30~dB. The minimum distance and the reference distance are both set to one meter $r_0=1m$, and the disc radius is set to $R=100 m$. The target rate is set to $R_m=1.5$ bits per channel use (BPCU). \textcolor[rgb]{0.00,0.00,1.00}{To mitigate the computational complexity, the number of Monte Carlo iterations is limited to $10^6$.}

\begin{figure}[t]
\centering
\includegraphics[width = 3in]{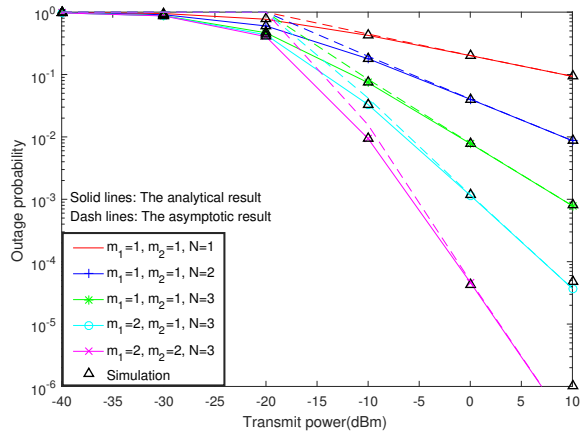}
\caption{\textcolor[rgb]{0.00,0.00,1.00}{OP of the $m$-th user, where the number of TAs/RAs are set to $M=K=1$. Both the BS-IRS and IRS-user links are NLoS, where $t_1=2, t_2=1$. The analytical results and asymptotic results are calculated from~\eqref{outage analytical results m-th in theorem2 minimum achivable} and~\eqref{asymptotic result m user in corollary1}, respectively.}}
\label{Outage LIS impact on N=1 fig 1}
\end{figure}

We first evaluate the OP impacted by the number of IRS elements in Fig.~\ref{Outage LIS impact on N=1 fig 1}. The solid and dashed curves represent the analytical results and asymptotic results, respectively. \textcolor[rgb]{0.00,0.00,1.00}{Since the results are derived as high-SNR approximations, the gaps between the simulation and analytical results are large. However, there is a close agreement between the simulation and analytical results in the high-SNR regime.}
Observe that the performance of OP increases as the number of IRS elements enhances, which is because that, as more IRS elements are employed, the channel gain is improved by the increased diversity order. For example, as shown by the blue curve and green curve, as well as by the results in~\eqref{diversity order}, the diversity orders are 4 and 6 for the cases of $N=2$ and $N=3$, respectively. Observe that the slope of curves increases with the number of IRS elements, which validates our \textbf{Remark~\ref{remark2:impact of antenna numbers}}.

\begin{figure}[t]
\centering
\includegraphics[width = 3in]{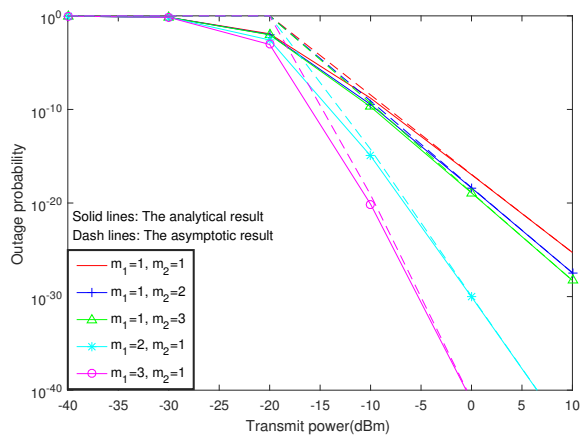}
\caption{ OP of the $m$-th user with $t_1, t_2=1, 2, 3$. The number of TAs and RAs, as well as IRS elements are set to $M=1, K=10$, $N=10$, respectively. The analytical results and asymptotic results are calculated based on the distribution from~\eqref{New Gamma distribution in Lemma}.}
\label{Outage LIS fading fig 2}
\end{figure}

We then evaluate the OP of the $m$-th user with different fading parameters in Fig.~\ref{Outage LIS fading fig 2}. We can see that as the SNR increases, the OP decreases. We can also observe that stronger LoS environments typically decrease the OP. Observe that the fading parameter of the IRS-user link has almost no effect on the OP, which mainly depends on that of the BS-IRS link. This phenomenon verifies the insights gleaned from \textbf{Remark~\ref{remark1:impact of fading environment}}.
Note that the slope of the curves is governed by $t_1$, which verifies that the diversity orders of the schemes mainly depend on $t_1$, when the effective channel gain is high enough. This also validates the insights from \textbf{Remark~\ref{remark2:impact of antenna numbers}}.

\begin{figure}[t!]
\centering
\includegraphics[width =3in]{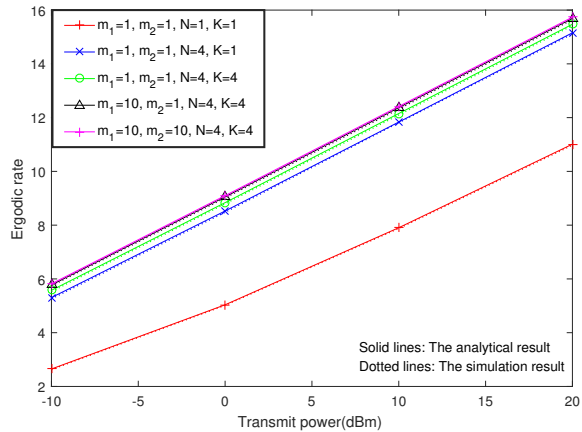}
\caption{Ergodic rate of the $m$-th user in the case of $M=1$.}
\label{ergodic rate LIS fading fig 5}
\end{figure}

\cbstart We evaluate the ergodic rate of the $m$-th user in Fig.~\ref{ergodic rate LIS fading fig 5}. The solid and dashed curves represent the analytical and simulation results, respectively. By comparing the black curve and cyan curve, one can observe that the ergodic rate of users is nearly not impacted by the LoS links between the IRSs and users, whereas the LoS links between the BS and IRSs improve the ergodic rate, which verifies~\textbf{Remark~\ref{remark6:impact of ergodic rate in massive LIS}}. In addition, the high-SNR slope of user $m$ is one, which illustrates that the number of IRS elements has no impact on the high-SNR slope of users. Furthermore, by employing more IRS elements, the ergodic rate can be significantly increased, which is because that spatial diversity gain can be significantly increased upon increasing the number of IRS elements.
\cbend

\begin{figure}[t!]
\centering
\includegraphics[width = 3in]{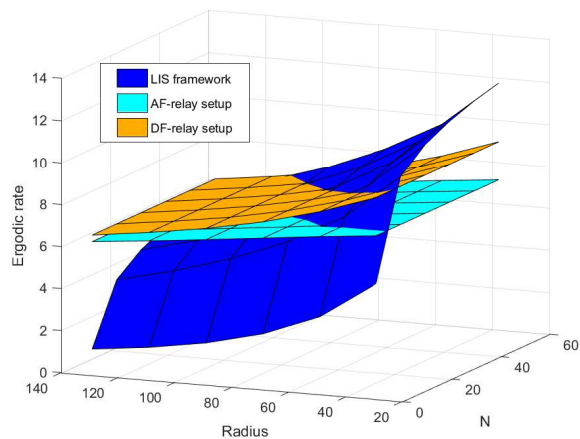}
\caption{\textcolor[rgb]{0.00,0.00,1.00}{ Spectral efficiency of the MIMO-IRS network, AF-relay and DF-relay setup versus the number of IRS elements and the total transmit power, where the number of TAs/RAs are set to $M=1, K=1$. The fading parameters are set to $t_1=3, t_2=1$.}}
\label{Compare With AF_DF relay fading fig 6}
\end{figure}

Combining with the insights from~\cite{DF_relaying_outage,renzo_RIS_relay}, the achievable rate of two alternative half-duplex relay setups, namely of the classic amplify-then-forward (AF) relay and of decode-then-forward (DF) relay setup are evaluated, where the transmission is divided into two equal-duration phases. It is assumed that the BS, relay and user are equipped with a single antenna in both the AF-relay and DF-relay setups.
We first evaluate the achievable rate of the AF-relay setup, where the AF-relay simply amplifies the received signal without decoding.
Therefore, on the one hand, the SE of the AF-relay setup can be written as
\begin{equation}\label{AF_relay_expectation}
R_{\rm AF} = \mathbb{ E} \left\{ {\frac{1}{2}{{\log }_2}\left( {1 + SIN{R_{{\rm{AF}}}}} \right)} \right\},
\end{equation}
where $p_d$ denotes the transmit power of the relay, $SIN{R_{{\rm{AF}}}} = \frac{{{\varepsilon _a}{{({d_1}{d_{2,m}})}^{ - \alpha }}{{\left| {{h_1}} \right|}^2}{{\left| {{h_2}} \right|}^2}}}{{{\sigma ^2}(1 + {\varepsilon _a}d_{2,m}^{ - \alpha }{{\left| {{h_2}} \right|}^2})}}$, and ${\varepsilon _a}$ denotes the amplification coefficient of the AF-relay with ${\varepsilon _a} = \frac{{{p_d}}}{{{p_b}}}d_1^\alpha {\left| {{h_1}} \right|^{ - 2}}$. Note that the AWGN of the BS to IRS link is also amplified by the AF-relay.

On the other hand, the achievable rate expression of the DF-relay setup is more complicated. The DF-relay has to decode the signal received from the BS, and the achievable rate can be written as follows:
\begin{equation}\label{DF_relay decode signal}
{R_{{\rm{DF,1}}}} = \mathbb{ E}\left\{ {\frac{1}{2}{{\log }_2}\left( {1 + SN{R_{{\rm{DF1}}}}} \right)} \right\},
\end{equation}
where $SN{R_{{\rm{DF1}}}} = \frac{{{p_b}d_1^{ - \alpha }{{\left| {{h_1}} \right|}^2}}}{{{\sigma ^2}}}$. Then, the user decodes the signal retransmitted by the DF-relay at a rate of:
\begin{equation}\label{DF_relay_user}
{R_{{\rm{DF,2}}}} = \mathbb{E}\left\{ {\frac{1}{2}{{\log }_2}\left( {1 + SN{R_{{\rm{DF2}}}}} \right)} \right\},
\end{equation}
where $SN{R_{{\rm{DF2}}}} = \frac{{{p_d}d_2^{ - \alpha }{{\left| {{h_2}} \right|}^2}}}{{{\sigma ^2}}}$.
Thus, the achievable rate of the DF-relay setup can be rewritten as
\begin{equation}\label{DF relay}
\begin{aligned}
{R_{{\rm{DF}}}} = {\min}\{ {R_{{\rm{DF,1}}}},{R_{{\rm{DF,2}}}}\}.
\end{aligned}
\end{equation}

\textcolor[rgb]{0.00,0.00,1.00}{ Here, we evaluate the SE of both our IRS network, as well as that of the AF-relay and DF-relay setups in Fig.~\ref{Compare With AF_DF relay fading fig 6}.}
The results of the AF-relay and DF-relay setup are derived from~\eqref{AF_relay_expectation} and~\eqref{DF relay}, respectively. \textcolor[rgb]{0.00,0.00,1.00}{ In order to compare our IRS network, as well as the AF-relay and DF-relay setups under fair conditions, the total transmit powers $p_{tot}$ are set to be identical. Explicitly, the transmit power of the IRS network is set to $p_b=p_{tot}$ dBm, while the transmit powers of both the AF-relay and DF-relay are set to $(p_d + p_b)=p_{tot}$ dBm. The optimal power split strategies of both the AF and DF relays are based on~\cite{renzo_RIS_relay} for achieving the optimal SE.} We can see that the rate difference between the MIMO-IRS network, AF-relay and DF-relay setup becomes smaller, when the number of IRS elements is increased. Observe that for the case of $N=15$, the proposed MIMO-IRS network is capable of outperforming the AF-relay and DF-relay setups, which indicates that the MIMO-IRS network becomes more competitive, when the number of IRS elements is high enough.
\textcolor[rgb]{0.00,0.00,1.00}{ One can also observe that the proposed MIMO-IRS network is more sensitive to the total transmit power, which is due to the fact that at the current state-of-the-art IRSs cannot amplify the received signals.}

\begin{figure}[t!]
\centering
\includegraphics[width = 3in]{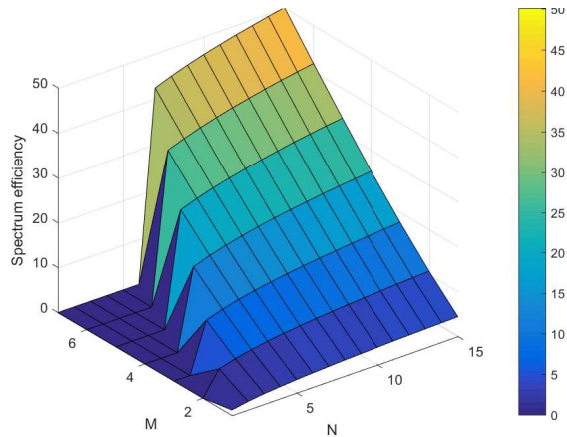}
\caption{Network throughput versus the number of TAs and RAs, as well as IRS elements with $K=M$. The LoS links are employed with $t_1=t_2=2$. }
\label{SE_LIS 3D fig 7}
\end{figure}

\cbstart \textcolor[rgb]{0.00,0.00,1.00}{ The network throughput is illustrated in Fig.~\ref{SE_LIS 3D fig 7}, where the transmit power is fixed to 30 dBm. One can observe that the network throughput increases with the number of TAs. This is due to the fact that the BS can simultaneously serve more users, hence providing a higher SE. Furthermore, the network's throughput monotonically increases as the number of TAs and of the number of IRS elements increases. Note that there is no solution for passive beamforming weights at the IRSs in the case of $N<MK$, which demonstrates that the network's SE is zero.} \cbend

\begin{figure}[t!]
\centering
\includegraphics[width = 3in]{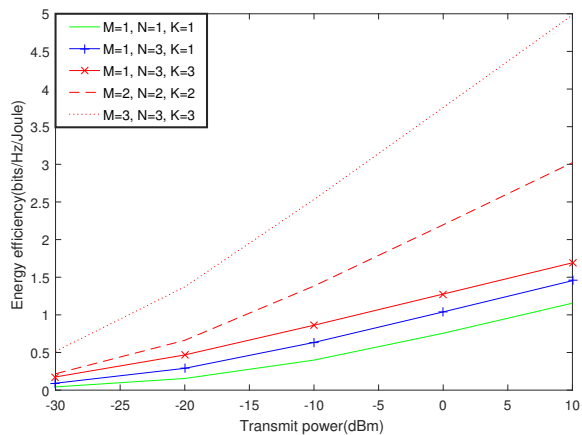}
\caption{EE of the proposed MIMO-IRS network with $t_1=5, t_2=1$. The amplifier efficiency at the BS is set to ${\varepsilon _b}$ =1.2. The static power at the BS and users are set to $P_{\rm{B,s}}=9$ dBW and $P_{\rm{U}}=10$ dBm, respectively. The power consumption for a single IRS element is set to $P_L=10$ dBm. }
\label{EE_LIS fig 8}
\end{figure}

\cbstart The network's EE is evaluated in Fig.~\ref{EE_LIS fig 8}. Observe that the network's EE improves as the number of RAs, TAs and IRS elements increases, which is because that increasing the number of RAs, TAs and IRS elements is capable of increasing the spatial diversity gain. Moreover, the slope of EE can be improved by improving the number of RAs, TAs and IRS elements. \cbend

\section{Conclusions}
\cbstart
We first discussed the recent developments in the IRS network.
We then designed the detection vectors of the users and the passive beamforming at the IRSs, where the proposed network performs well both in IRS and non-IRS scenarios. Furthermore, we derived the closed-form of channel characteristics, outage probabilities, ergodic rates, SEs and EEs.
\textcolor[rgb]{0.00,0.00,1.00}{In NG networks, multiple IRS arrays could be deployed to the infrastructures and other local paraphernalia. Therefore, one promising future direction is distributed IRS network, where multiple IRS arrays are beneficially distributed on the facades of different buildings located in different areas.} \textcolor[rgb]{0.00,0.00,1.00}{Based on the insights from~\cite{reconfig_meta_surf_2,stochastic_random}, another important future direction of IRS and stochastic geometry  research is to model the location of users in a restricted geographic area.} \cbend

\numberwithin{equation}{section}
\section*{Appendix~A: Proof of Lemma~\ref{lemma2:new state of effective channel gain exact high SNR for OP}} \label{Appendix:As}
\renewcommand{\theequation}{A.\arabic{equation}}
\setcounter{equation}{0}

\cbstart Since the elements of ${\left| {{{\rm \bf{G}}_{m}}} \right|}$ and ${\left| {{{\rm \bf{H}}}} \right|}$ are i.i.d., and according to~\cite{Naka_multiple}, the PDF of the product of two Nakagami-$m$ random variables can be given by \cbend
\begin{equation}\label{Apeendix effective channel gain_A_first}
\begin{aligned}
{f_{\left | {{{{\rm{ h}}}_m}} \right |^2}}(x) & = \frac{{4{{\left( {{t_s}{t_l}} \right)}^{\frac{{{t_s} + {t_l}}}{2}}}}}{{\Gamma \left( {{t_s}} \right)\Gamma \left( {{t_l}} \right)}}{x^{{t_s} + {t_l} - 1}}\\
& \times {I_{{t_s} - {t_l}}}\left( {2\sqrt {{t_s}{t_l}} x} \right),
\end{aligned}
\end{equation}
where ${I_{\alpha}}\left( \cdot \right)$ denotes the modified Bessel function of the second kind. \textcolor[rgb]{0.00,0.00,1.00}{ Since~\eqref{Apeendix effective channel gain_A_first} contains the modified Bessel function of the second kind, which is hard to calculate, hence we use the Laplace transform of the PDF, which can be derived as }

\begin{equation}\label{appendix A channel laplace transform}
\begin{aligned}
& {L_{\left | {{{{\rm \bf{ h}}}_m}} \right|^2}}(s) = { \mathbb E} \left( {{e^{ - s\left\| {{{{\rm{\hat h}}}_m}} \right\|_2^2}}} \right) = \frac{{4{{\left( {{t_s}{t_l}} \right)}^{\frac{{{t_s} + {t_l}}}{2}}}}}{{\Gamma \left( {{t_s}} \right)\Gamma \left( {{t_l}} \right)}}\\
& \times \int_0^\infty  {{x^{{t_s} + {t_l} - 1}}{e^{ - sx}}{I_{{t_s} - {t_l}}}\left( {2\sqrt {{t_s}{t_l}} x} \right)dx}.
\end{aligned}
\end{equation}

By utilizing~\cite[eq. (6.621.3)]{Table_of_integrals},~\eqref{appendix A channel laplace transform} can be further transformed into
\begin{equation}\label{appendix A first laplace tranform expression}
\begin{aligned}
 & {L_{\left | {{{{\rm \bf{ h}}}_m}} \right|^2}}(s)= {\bar m}{\left( {s + 2\sqrt {{t_s}{t_l}} } \right)^{ - 2{t_s}}}\\
& \times F\left( {2{t_s},{t_s} - {t_l} + 0.5;{t_s} + {t_l} + 0.5;\frac{{s - 2\sqrt {{t_s}{t_l}} }}{{s + 2\sqrt {{t_s}{t_l}} }}} \right),
\end{aligned}
\end{equation}
where ${\bar m} = \frac{{\sqrt \pi  {4^{{t_s} - {t_l} + 1}}{{\left( {{t_s}{t_l}} \right)}^{  {t_s}}}\Gamma \left( {2{t_s}} \right)\Gamma \left( {2{t_l}} \right)}}{{\Gamma \left( {{t_s}} \right)\Gamma \left( {{t_l}} \right)\Gamma \left( {{t_s} + {t_l} + 0.5} \right)}}$, and $F\left( { \cdot, \cdot ; \cdot ;\cdot } \right)$ represents the hypergeometric series.

\textcolor[rgb]{0.00,0.00,1.00}{ We then focus on the high-SNR regime, where we have ${\frac{{{P_b}}}{{{\sigma ^2}}} \to \infty }$ and $s \to \infty $. Then, upon assuming ${t_1} \ne {t_2}$, the Laplace transform can be rewritten as}
\begin{equation}\label{appendix A laplace transform high SNR regime}
\begin{aligned}
{L_{\left | {{{{\rm \bf{ h}}}_m}} \right |^2}}(s) = \tilde m{\left( {s + 2\sqrt {{t_s}{t_l}} } \right)^{ - 2{t_s}}}.
\end{aligned}
\end{equation}

Since the reflected signals are i.i.d., the overall effective channel gain can be given by
\begin{equation}\label{appendix A effective channel distribution before inverse}
{L_{\left | {{{{\rm{\hat h}}}_m}} \right |^2}}(s) = {{\tilde m}^{N}}{\left( {s + 2\sqrt {{t_s}{t_l}} } \right)^{ - 2{t_s}N}}.
\end{equation}
Based on the inverse Laplace transform for~\eqref{appendix A effective channel distribution before inverse}, the PDF and CDF in the high-SNR regime can be obtained in~\eqref{New Gamma distribution in Lemma}, and hence we complete the proof.

\numberwithin{equation}{section}
\section*{Appendix~B: Proof of Theorem~\ref{Theorem1:Outage m-th user closed form by hypergeom}} \label{Appendix:Bs}
\renewcommand{\theequation}{B.\arabic{equation}}
\setcounter{equation}{0}

We first handle the OP defined of user $m$ in~\eqref{Outage Defination}, which can be rewritten as
\begin{equation}\label{Appendix C outage probability expectation}
{P_{m,l}} =\mathbb{ P}\left( {\left | {{{{\rm \bf{{{\rm{ h}}}}}}_m}} \right |^2 < {\delta _{m,l}}{{({d_1}  {d_{2,m}})}^\alpha }} \right).
\end{equation}

\textcolor[rgb]{0.00,0.00,1.00}{ Then, based on the channel statistics derived in~\eqref{New Gamma distribution in Lemma_high-SNR-exact PDF} and on the distance distribution in~\eqref{PDF of distance distribution}, the OP can be transformed into}
\begin{equation}\label{appendix C first outage expression}
\begin{aligned}
{P_{m,l}} &= \frac{{2{{\tilde m}^{N}}{{\left( {4{t_s}{t_l}} \right)}^{ - {t_s}N}}}}{{\Gamma \left( {2{t_s}N} \right)\left( {{R^2} - r_0^2} \right)}}\\
&\times \int\limits_{{r_0}}^R {\gamma \left( {2{t_s}N,2\sqrt {{t_s}{t_l}} {\delta _{m,l}}d_1^\alpha {r^\alpha }} \right)} rdr.
\end{aligned}
\end{equation}
By some further algebraic manipulations, we arrive at:
\begin{equation}\label{appendix C ourage transformed but before integral}
{P_{m,l}}\mathop  = \limits^{(c)} \varphi {b^{ - {\delta _1}}}\int\limits_{br_0^\alpha }^{b{R^\alpha }} {\gamma \left( {a,bx} \right)} {x^{{\delta _1}}}dx,
\end{equation}
where $(c)$ is obtained by substituting $x ={b}{r^\alpha }$.

Then, based on~\cite{Table_of_integrals}, the OP can be calculated as
\begin{equation}\label{Appendix C J1_expression}
\begin{aligned}
{P_{m,l}} &= {\frac{{\varphi {b^a}{R^{\alpha a + 2}}}}{{a(\alpha a + 2)}}}{}_2{F_2}\left( {a,a + {\delta _2};a + 1,a + {\delta _2} + 1; - b{R^\alpha }} \right) \\
& - {\frac{{\varphi {b^a}r_0^{\alpha a + 2}}}{{a(\alpha a + 2)}}}{}_2{F_2}\left( {a,a + {\delta _2};a + 1,a + {\delta _2} + 1; - br_0^\alpha } \right).
\end{aligned}
\end{equation}
Thus, the closed-form OP expression can be obtained as in~\eqref{outage analytical results m-th in theorem1}, and the proof is complete.

\numberwithin{equation}{section}
\section*{Appendix~C: Proof of Corollary~\ref{corollary1:Outage m user asymptotic}} \label{Appendix:Cs}
\renewcommand{\theequation}{C.\arabic{equation}}
\setcounter{equation}{0}

For the case of $bR^\alpha  < 1$, the hypergeometric function can be expanded to
\begin{equation}\label{Appendix D hypergeo expansion}
\begin{aligned}
&{}_2{F_2}\left( {a,a + {\delta _1};a + 1,a + {\delta _1} + 1; - bR^\alpha } \right) \\
& = \sum\limits_{n = 0}^\infty  {\left( {\frac{{{{(a)}_n}{{(a + {\delta _1})}_n}}}{{{{(a + 1)}_n}{{(a + {\delta _1} + 1)}_n}n!}}} \right)} {b^n}R^{\alpha n}.
\end{aligned}
\end{equation}

Thus, we can readily arrive at the asymptotic result of
\begin{equation}\label{Appendix D J1 asymptotic}
{{\bar P}_m} = \frac{\varphi }{{a(\alpha a + 2)}}\sum\limits_{n = 0}^\infty  {{t_{n1}}} {b^{n + a}}\left( {R^{\alpha a + \alpha n + 2} - r_{0 }^{\alpha a + \alpha n + 2}} \right).
\end{equation}
We then can derive the desired result by algebraic manipulations in~\eqref{asymptotic result m user in corollary1}. The proof is complete.

\numberwithin{equation}{section}
\section*{Appendix~D: Proof of Lemma~\ref{lemma2:new state of effective channel gain exact}} \label{Appendix:Ds}
\renewcommand{\theequation}{D.\arabic{equation}}
\setcounter{equation}{0}

Since the elements of ${\left| {{{\rm \bf{G}}_{m}}} \right|}$ and ${\left| {{{\rm \bf{H}}}} \right|}$ are i.i.d., we have the following inequations of the $m$-th user as
\textcolor[rgb]{0.00,0.00,1.00}{ \begin{equation}\label{Apeendix effective channel gain}
\begin{aligned}
{\left | {{{{\rm \bf{ h}}}_m}} \right |^2} &= \left( \sum\limits_{k = 1}^K {} \sum\limits_{n = 1}^{N} {} {\left| {{g_{m,k,n}}{h_{n,m}}} \right|} \right)^2\\
& \ge \sum\limits_{k = 1}^K {} \sum\limits_{n = 1}^{N} {} {\left| {{g_{m,k,n}}} \right|^2}{\left| {{h_{n,m}}} \right|^2}.
\end{aligned}
\end{equation}}

Then, based on the property of Nakagami distribution, we can have:
\begin{equation}\label{appendix channel multiple distribution}
\left( {\sum\limits_{k = 1}^K {{{\left| {{ g_{m,k,n}}} \right|}^2}} } \right) \sim \mathcal{RV}\left( {K,{{\frac{K}{{{t_2}}}}}} \right).
\end{equation}
\textcolor[rgb]{0.00,0.00,1.00}{ Then the mean and variance of the effective channel gain in the case of $N=1$ can be expressed as
\begin{equation}\label{appendix first mean}
\mathbb{ E}\left( {\sum\limits_{k = 1}^K {{{\left| {{g_{m,k,n}}} \right|}^2}} } \right) \mathbb{E}\left( {{{\left| {{{\rm{h}}_{1,m}}} \right|}^2}} \right) = K,
\end{equation}}
and
\begin{equation}\label{appendix first variance}
\begin{aligned}
{V}  = \frac{{K\left( {1 + {t_1} + K{t_2}} \right)}}{{{t_1}{t_2}}}.
\end{aligned}
\end{equation}

Hence, by utilizing zero-forcing based detection vectors, the mean and variance of the channel gain can be obtained as
\begin{equation}\label{appendix effective channel distribution}
\left | {{{{{{\rm \bf{ h}}}}}_m}} \right |^2 \sim \mathcal{RV} \left( {NQ,NQ{t_h} } \right),
\end{equation}
where ${t_h}=\frac{{\left( {1 + {t_1} + Q{t_2}} \right)}}{{{t_1}{t_2}}}$.
Hence, the proof is complete.

\numberwithin{equation}{section}
\section*{Appendix~E: Proof of Theorem~\ref{theorem3:ergodic rate m-th user}} \label{Appendix:Es}
\renewcommand{\theequation}{E.\arabic{equation}}
\setcounter{equation}{0}

We initiate by providing the definition of ergodic rate as follows:
\begin{equation}\label{Appendix E first define}
\begin{aligned}
{R_{m,e}} & = \mathbb{E} \left\{ {{{\log }_2}\left( {1 + SN{R_m}\left( x \right)} \right)} \right\}\\
& =  - \int\limits_0^\infty  {{{\log }_2}(1 + x)} d\left( {1 - {{F}}\left( x \right)} \right) \\
& = \frac{1}{{\ln \left( 2 \right)}}\int\limits_0^\infty  {\frac{{1 - F\left( x \right)}}{{1 + x}}} dx.
\end{aligned}
\end{equation}

By utilizing the results of CDF, we set
\begin{equation}\label{Appendix E CDF define}
\lambda= 1-F\left( x \right) = \varphi \int\limits_{{r_0}}^R {\Gamma_u \left( {a,cx{{r}^\alpha }} \right)} rdr,
\end{equation}
where $\Gamma_u (,)$ denotes the upper incomplete Gamma function.

By substituting $t = cx{r^\alpha }$ into~\eqref{Appendix E CDF define}, $\lambda$ can be transformed into
\begin{equation}\label{Appendix E ergodic before second expansion}
\begin{aligned}
\lambda = {\frac{{\varphi {{(cx)}^{ - {\delta _2}}}}}{\alpha }\int\limits_{cx{r_0^\alpha}}^{cx{R^{\alpha}}} {\Gamma_u \left( {a,t} \right)} {t^{ {\delta _2} - 1}}dt} .
\end{aligned}
\end{equation}

By using upper incomplete Gamma function, $\lambda$ is derived as
\begin{equation}\label{Appendix E J3 first}
\begin{aligned}
\lambda &= \frac{\varphi }{2}R^2\Gamma_u \left( {a,cR^\alpha x} \right) - \frac{\varphi }{2}{(cx)^{ - {\delta _2}}}\Gamma_u \left( {a + {\delta _2},cR^\alpha x} \right)\\
& - \frac{\varphi }{2}r_{0}^2\Gamma_u \left( {a,cr_{0 }^\alpha x} \right) + \frac{\varphi }{2}{(cx)^{ - {\delta _2}}}\Gamma_u \left( {a + {\delta _2},cr_{0 }^\alpha x} \right).
\end{aligned}
\end{equation}

To obtain tractable ergodic rates, the upper incomplete Gamma function are transformed into the Meijer-G function~\cite{Table_of_integrals} as
\begin{equation}\label{appendix E upper gamma expand to meijer G}
\Gamma_u \left( {a,cR^\alpha x} \right) = {\rm{G}}
\begin{tiny}
\begin{array}{*{20}{c}}
{2,0}\\
{1,2}
\end{array}
\end{tiny}
\left( {\left. {\begin{array}{*{20}{c}}
1\\
{a,0}
\end{array}} \right|cR^\alpha x} \right).
\end{equation}
Then, based on \cite{Table_of_integrals}, the ergodic rate of the $m$-th user is obtained as
\begin{equation}\label{Appendix E J3 final}
\begin{aligned}
\int\limits_0^\infty  {\frac{\lambda}{{1 + x}}} dx &= \frac{\varphi }{2}R^2{\rm{G}}
\begin{tiny}
\begin{array}{*{20}{c}}
{3,1}\\
{2,3}
\end{array}
\end{tiny}
\left( {\left. {\begin{array}{*{20}{c}}
{0,1}\\
{0,0,a}
\end{array}} \right|cR^\alpha } \right) \\
& - \frac{\varphi }{2}{c^{ - {\delta _2}}}{\rm{G}}
\begin{tiny}
\begin{array}{*{20}{c}}
{3,1}\\
{2,3}
\end{array}
\end{tiny}
\left( {\left. {\begin{array}{*{20}{c}}
{{\delta _2},1}\\
{{\delta _2},0,a + {\delta _2}}
\end{array}} \right|cR^\alpha } \right) \\
& - \frac{\varphi }{2}r_{0 }^2{\rm{G}}
\begin{tiny}
\begin{array}{*{20}{c}}
{3,1}\\
{2,3}
\end{array}
\end{tiny}
\left( {\left. {\begin{array}{*{20}{c}}
{0,1}\\
{0,0,a}
\end{array}} \right|cr_{0 }^\alpha } \right) \\
& + \frac{\varphi }{2}{c^{ - {\delta _2}}}{\rm{G}}
\begin{tiny}
\begin{array}{*{20}{c}}
{3,1}\\
{2,3}
\end{array}
\end{tiny}
\left( {\left. {\begin{array}{*{20}{c}}
{{\delta _2},1}\\
{{\delta _2},0,a + {\delta _2}}
\end{array}} \right|cr_{0 }^\alpha } \right).
\end{aligned}
\end{equation}
Thus, we can derive the ergodic rate in~\eqref{asympto m-th erogodic rate in corollary4}, which completes the proof.

\bibliographystyle{IEEEtran}
\bibliography{IEEEabrv,IS_SG}

\end{document}